\documentclass[12pt,a4paper]{article}
\usepackage{epsfig}
\newcommand{\be}{\begin{equation}}
\newcommand{\ee}{\end{equation}}
\newcommand{\bea}{\begin{eqnarray}}

\newcommand{\eea}{\end{eqnarray}}
\newcommand{\nn}{\nonumber}
\newcommand{\dd}{\displaystyle}
\newcommand{\spur}[1]{\not\! #1 \,}
\def\slash#1{\setbox0=\hbox{$#1$}#1\hskip-\wd0\dimen0=5pt\advance
\dimen0 by-\ht0\advance\dimen0 by\dp0\lower0.5\dimen0\hbox
to\wd0{\hss\sl/\/\hss}}
\def\Black{}
\def\Blue{}
\def\Brown{}
\begin{document}
\begin{titlepage}
\title{\hfill $\mbox{\small{\begin{tabular}{r} $\Blue{\rm
Bari-TH/99-361}$\\ $\Blue{\rm Napoli-DSF~99/32}$
\end{tabular}}}$ \\[1truecm]
\Brown $B\to \pi \pi \ell \nu_\ell$ decays \\ in a QCD
relativistic potential model}

\vspace{+1truecm}

\author{\Black M. Ladisa$^{a,b}$, G. Nardulli$^{a}$, T. N. Pham$^{b}$,
P. Santorelli$^{c}$}

\date{~}
\maketitle

\vspace{-2truecm}
\begin{it}
\begin{center}
$^a$Dipartimento di Fisica dell'Universit\`a di Bari, Italy\\
Istituto Nazionale di Fisica Nucleare, Sezione di Bari,
Italy\\\vspace*{0.3cm}  $^b$ Centre de Physique Th\'eorique, \\
Centre National de la Recherche Scientifique, UMR 7644, \\ \'Ecole
Polytechnique, 91128 Palaiseau Cedex, France\\  \vspace*{0.3cm}
$^c$Dipartimento di Scienze Fisiche, Universit\`a "Federico II" di
Napoli, Italy\\Istituto Nazionale di Fisica Nucleare, Sezione di
Napoli, Italy\vspace*{0.3cm}
\end{center}
\end{it}
\begin{abstract}
\noindent In the framework of a QCD relativistic potential model
we evaluate the form factors describing  the exclusive decay $B
\to \pi\pi \ell \nu$. The calculation is performed in a phase
space region far away from  the resonances and therefore is
complementary to other decay mechanisms where the pions are
produced by intermediate particles, e.g. in the chiral approach.
We give an estimate of the contribution of the non resonant
channel of the order of $\mathcal{B}$$(B^- \to \pi^+ \pi^- \ell^-
\bar \nu_\ell)\simeq 2.2 \times 10^{-4}$.
\end{abstract}\thispagestyle{empty}
\end{titlepage}
\setcounter{page}{1}

In this letter we shall study the $B-$meson decays \bea && B^- \to
\pi^+\pi^- \ell^- \bar \nu_\ell \label{semilep1}
\\
&&\bar B^0 \to \pi^+\pi^0  \ell^- \bar \nu_\ell~. \label{semilep2}
\eea

>From the experimental side these decays are interesting in view of
the future programs at the $B$-factories. For example, some of the
preliminary studies on  the CP violations at these machines
\cite{babar} have examined the possibility to extract the angle
$\alpha$ of the unitarity triangle by the $B\to\rho\pi$ non
leptonic decay channel. The non-resonant decay mode $B\to\ 3 \pi$
would be interesting to analyze in this context, as it might
provide a significant background to the main decay process. While
a calculation from first principles is not available at the
moment, a useful approximation might be the factorization
approximation and, within this approximation, the decay modes
(\ref{semilep1}) and (\ref{semilep2}) would provide the crucial
hadronic matrix elements needed to compute the relevant
amplitudes. In passing we note that there is another channel, {\it
i.e.} the $ B^- \to \pi^0\pi^0 \ell^- \bar \nu_\ell $ decay mode,
which will not be examined here because it is less interesting
from an experimental point of view.

>From a theoretical standpoint semileptonic $B$-meson decays with
two hadrons in the final state represent a formidable challenge as
they involve hadronic matrix elements of weak currents with three
hadrons. They can be studied by pole diagrams, which amounts to a
simplification because only two hadrons are involved in the
hadronic matrix elements. This is the approach followed in some
papers where these decays have been examined  in the framework of
the chiral perturbation theories for heavy meson decays
\cite{lee},\cite{burdman}. This method is based on an effective
theory implementing both heavy-quark and chiral symmetry
\cite{burdman}, \cite{wise}, \cite{Wolfenstein}, \cite{casalbuoni}
and allows to achieve, for systems comprising both heavy ($Q$) and
light ($ q$) quarks, rigorous results  in the combined
$m_Q\to\infty,~m_q\to 0$ limit. However the range of validity of
this approach is limited by the requirement of soft pion momenta.
In the soft pion limit the amplitude is dominated by a few tree
diagrams with resonances as intermediate states, and some  clear
predictions can be made, but, at least for $B$ decays, the actual
phase space is relatively large  and the phenomenological interest
of these predictions is modest. The aim of this letter is to
examine the decays (\ref{semilep1}) and (\ref{semilep2}) in the
framework of a QCD relativistic potential model \cite{pietroni}
and to extend the kinematical range where theoretical predictions
are possible. We shall present a detailed analysis of the four
form factors relevant to (\ref{semilep1}); for reasons of space we
shall only give a prediction for the width of the decay
(\ref{semilep2}). We shall not include  final state interactions
in our calculation since no consistent way to compute them is
presently available. It's clear however that they can modify our
numerical calculations \cite{nard-pham}.

\par
In two recent papers: \cite{Brho} ,\cite{Bpi} we have presented an
analysis of some semileptonic and rare  $B$ decays into one light
hadron employing the relativistic potential model in an
approximation that renders the calculations simpler. We wish to
exploit here this approximation in the study of the $B\to \pi\pi
\ell  \nu$ decays .

Let us start with a description of the model (for more details see
\cite{pietroni}, \cite{Brho}  and \cite{Bpi}). In this approach
the mesons are described as bound states of constituent quarks and
antiquarks tied by an instantaneous potential $V(r)$, which  has a
confining linear behaviour at large interquark distances $r$ and a
Coulombic behaviour $\simeq -\alpha_s (r)/r$ at small distances,
with $\alpha_s (r)$ the running strong coupling constant (the
Richardson's potential \cite{Rich}  is used to interpolate between
the two regions\footnote{Spin terms  are not included in $V(r)$,
which, for heavy mesons, is justified by the spin symmetry  in the
limit $m_Q \to \infty$. Their neglect cannot be justified for
light mesons, which is one of  the reasons why one does not use
the constituent quark picture for the pions.}). Due to the nature
of the interquark forces, the light quarks are
relativistic; for this reason one employs for the meson wave
function $\Psi$ the Salpeter \cite{salpeter} equation embodying
the relativistic kinematics:
\be
\left [ \sqrt{ -\nabla_1^2 + m^2_1} + \sqrt{ -\nabla_2^2 + m^2_2}
+V(r)\right ] \Psi(\vec r)=M \Psi(\vec r)~ \label{salp} \;\; , \ee
where the index 1 refers to the heavy quark and the index 2 to the
light antiquark; $M$ is the heavy meson mass that is obtained by
fitting the various parameters of the model, in particular the
b-quark mass, that is fitted to the value $m_b=4890$ MeV, and the
light quark masses $m_u\simeq m_d=38$ MeV, $m_s=115$ MeV.
 The $B$-meson wave function $\Psi(\vec r)$ in its rest
frame is obtained by solving eq. (\ref{salp}); a useful
representation in Fourier momentum space was obtained in \cite{Brho}
and is as follows
\be
\psi (k)=4\pi\sqrt{m_B \alpha^3}\ e^{-\alpha k} \;\; , \label{wf}
\ee with $\alpha=2.4$ GeV$^{-1}$ and $k=|\vec k|$ the quark
momentum in the $B$ rest frame; this is the first approximation
introduced in \cite{Brho}.

The constituent quark picture used in the model is rather crude.
There are no propagating gluons in the instantaneous
approximation~: the Coulombic interaction is assumed to be static.
Moreover, the complex structure of the hadronic vacuum is
simplified: the confinement can be introduced by the linearly
rising potential at large distances, but the chiral symmetry and
the Nambu-Goldstone boson nature of the $\pi$'s cannot be
implemented by the constituent quark picture. For these reasons,
while there are good reasons to believe that eq. (\ref{salp}) may
describe the quark distribution inside the heavy meson, one cannot
pretend to apply it to light mesons. Therefore pion couplings to
the quark degrees of freedom are described by effective vertices.

To evaluate the amplitude for semileptonic decays, it is useful to
follow some simple rules, similar to the Feynman rules by which
the amplitudes are computed in perturbative field theory. The
setting of these rules is the main innovation introduced in
\cite{Brho} as compared to \cite{pietroni}. For the decays
(\ref{semilep1}) and (\ref{semilep2}) we draw a quark-meson
diagram as in fig. \ref{diagram} and we evaluate it according to
the following rules:
\par\noindent
1) for a charged pion  of momentum $p_{\pi}$ we write the coupling
\be
-\frac{N_q\ N_{q^{\prime}}}{f_\pi} \spur{p_{\pi}} \gamma_5 \;\;,
\label{0-} \ee where $f_{\pi}=130\ MeV$. The normalization factors
$N_q,~ N_{q^{\prime}}$ for the quark coupled to the meson are
discussed below;\\
 \noindent
2) for the heavy meson  $B$ in the initial state one introduces
the matrix:
 \be
 B=\frac{1}{\sqrt3}\psi (k){\sqrt{ \frac{m_q m_b} {m_q m_b+q_1\cdot
 q_2} }}\;\;\frac{\spur{q_1}+m_b}{2 m_b}(-i\ \gamma_5)
 \frac{-\!\!\!\spur{q_2}+m_q}{2 m_q} \label{B}
 \ee
where $m_b$ and $m_q$ are the heavy and light quark masses,
$q^\mu_1,\ q^\mu_2$ their $4-$momenta. The normalization factor
corresponds to the normalization $<B|B>=2\ m_B$ and $\dd\int
\frac{d^3 k}{(2\pi)^3} |\psi(k)|^2=2 m_B$ already embodied in
(\ref{B}). One assumes that the $4-$momentum is conserved at the
vertex $B\bar qb$, i.e. $q^\mu_1+q^\mu_2=p^\mu=$ $B$-meson
$4-$momentum. Therefore $q^\mu_1=(E_b,\vec k),~q^\mu_2=(E_q,-\vec
k)$ and
\be
E_b+E_q=m_B \;\; ; \label{Alt-Cab} \ee
\noindent
3) to take into account the
off-shell effects due to the quarks interacting in the meson, one
introduces running quark mass $m_b(k)$, to enforce the condition
\be
E=\sqrt{m^2(k)+|\vec k|^2} \label{Alt-Cab2} \ee for the
constituent quarks  \footnote{By this choice, the average
$<m_b(k)>$ does not differ significantly from the value $m_b$
fitted from the spectrum, see \cite{Brho} for details.};
\par
\noindent
4) the condition $m_b^2 \geq 0$ implies the constraint
\be
0\leq k\leq k_M\simeq \frac{m_B}{2 }~,\label{kmax} \ee on the
integration over the loop momentum $k$
\be
\int\frac{d^3k}{(2\pi)^3}\;\; ;
\label{loop}
\ee
\noindent
5) for each quark line with momentum $q$ and not representing a constituent
quark one introduces the factor
\be
\frac{i}{\spur q-m_{q^{\prime}}}\times G(q^2)~,
\label{ff}
\ee
where $G(q^2)$ is a shape function that modifies the free propagation of the
quark of mass $m_{q^{\prime}}$ in the hadronic matter. The shape function
\be
G(q^2)=\frac{m^2_G-m^2_{q^{\prime}}} {m^2_G-q^2} \label{ff2} \ee
was  adopted in \cite{Brho} and \cite{Bpi},  with the the value
$m_G^2=3~{\rm GeV}^2$ for  the mass parameter;
\par
\noindent
6) for the weak hadronic current one puts the factor
\be
N_{q} N_{q^\prime}\gamma^\mu (1-\gamma_5)  \;. \label{J} \ee
 The
normalization factor $N_q$ is as follows:
\be
N_q=
\left\{\begin{array}{ccl}
\dd\sqrt{\frac{m_q}{E_q}} &  ~~ &  {\rm(if ~q=constituent~ quark)} \\
 & \\
1 & ~~ & {\rm (otherwise) \; ;}
\end{array}\right .
\label{Nq} \ee \noindent 7) finally  the amplitude must contain a
colour factor of 3 and a trace over Dirac matrices; for the
$\pi^0$ coupling a further factor $\pm\frac{1}{\sqrt 2}$ is
introduced (the upper sign for coupling to the up quark, the lower
one for coupling to the down quark).

This set of rules can now be applied to the evaluation of the
hadronic matrix element for the decay (\ref{semilep1}),
corresponding to the diagram in fig. \ref{diagram}; the result
is~:
\begin{figure}[ht!]
\begin{center}
\epsfig{file=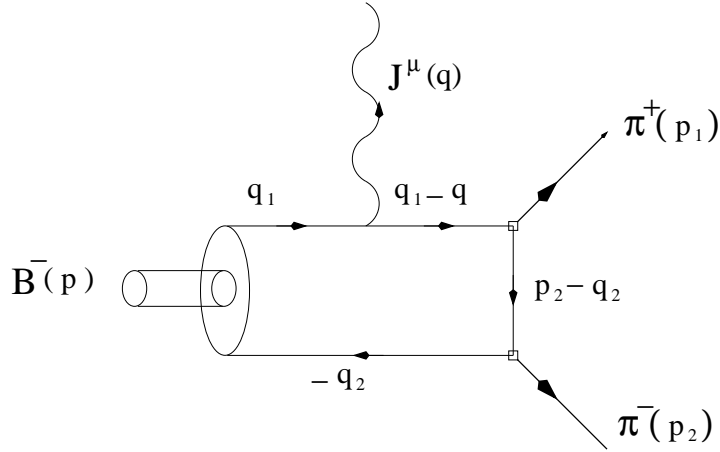,height=6cm}
\end{center}
\caption{The Feynman diagram for  $B^- \to \pi^+
\pi^-$ semileptonic decay.} \label{diagram}
\end{figure}

 \bea
 &&J^\mu=<\pi^+(p_{1})\pi^-(p_{2})|\bar u \gamma^\mu(1-\gamma_5)b|B^-(p)>
=\nn\\
 &&=
\frac{ i \sqrt{3}}{4 f^2_\pi} \int\frac{d^3k}{(2\pi)^3}~
 \frac{
\theta[k_M-k] ~\psi(k) }{ \sqrt{E_b E_q (m_q m_b+q_1\cdot q_2)}}
~\frac{G[(q_1-q)^2]}{(q_1-q)^2-m^2_{q^{\prime}}}\frac{G[(q_1-q-p_1)^2]}{(q_1-q-p_1)^2-
m^2_{q^{\prime\prime}}}\times \nn\\
 &&
 Tr\left[ (\spur{q_1}+m_b) (\spur{q_2}+ m_q) \spur{p_2}
 (\spur{p}_1+\spur{q}-\spur{q}_1-m_{q^{\prime\prime}}) \spur{p_1}
 (\spur{q}-\spur{q_1}-m_{q^{\prime}}) \gamma^{\mu}(1-\gamma_5)
 \right] \;.
 \label{ampiezza}
\eea
\noindent
The amplitude with a $\pi^0$ in the final state:
\begin{equation}\label{pionezero}
  J_0^\mu~=~<\pi^+(p_{1})\pi^0(p_{2})|\bar u \gamma^\mu(1-\gamma_5)b|\bar B^0(p)>
\end{equation}
is obtained from $ J^\mu(p_{1},~p_{2}, ~p)$ as follows:
\begin{equation}\label{pionezerobis}
  J_0^\mu(p_{1},~p_{2}, ~p)~=~-~\frac{1}{\sqrt{2}}~\Bigl [
 J^\mu(p_{1},~p_{2}, ~p)~-~J^\mu(p_{2},~p_{1}, ~p)
  \Bigr ]~.
\end{equation}

Following \cite{lee} we introduce the  various form factors as
follows. We put $q=p-p_1-p_2$ and we write
 \bea
 &&<\pi^+(p_{1})\pi^-(p_{2})|\bar u \gamma^\mu(1-\gamma_5)b|B(p)> =\nn\\
 &&=
i~w_+~(p_1+p_2)^{\mu}~+~i~w_-~(p_1-p_2)^{\mu}~+~{i}~r~q^\mu~+~
2~h~\epsilon^{\mu\alpha\beta\delta}~p_\alpha p_{1
\beta}p_{2\delta} \;\; . \label{Bpipi} \eea

\noindent It is useful to introduce the following variables:
\begin{eqnarray*}
  s~&=&~(p_1+p_2)^2 \\
t~&=&~(p-p_1)^2 \\ u~&=&~(p-p_2)^2~,
\end{eqnarray*}
that satisfy
\begin{equation}
s+t+u=q^2+m_B^2+2 m_{\pi}^2~.
\end{equation}

 The form factors
$h,~r,~w_-,~w_+$ are functions of three independent variables. One
can  choose as independent variables $s,~t~,q^2$ or,
alternatively, $s,~E_1,~E_2$, where $E_1,~E_2$ are the pion
energies in the $B$ rest frame. The relations between the two set
of invariants are:
\begin{eqnarray}\label{rel}
t&=&~m^2_B~+~m^2_\pi~-~2m_B~E_1 \nn\\
q^2&=&~s~+~m_B^2~-~2~m_B~(E_1~+~E_2)~.
\end{eqnarray}
Kinematical range is as follows:
\begin{eqnarray}\label{rel1}
4m_\pi^2 \leq & s & \leq m_B^2 \nn\\ 0 \leq & q^2 & \leq
(m_B-\sqrt{s})^2 \nn\\
\frac{m_B^2+2m_\pi^2+q^2-s}{2}-\frac{\lambda\sqrt{s-4m_\pi^2}}{2~\sqrt{s}}
\leq & t & \leq
\frac{m_B^2+2m_\pi^2+q^2-s}{2}+\frac{\lambda\sqrt{s-4m_\pi^2}}{2~\sqrt{s}},
\end{eqnarray}
where
\begin{equation}\label{p}
\lambda~=~\sqrt{(m^2_B-q^2+s)^2-4m^2_B~s}~.
\end{equation}

 From (\ref{ampiezza})  one can extract the different
 form factors by multiplying $J^\mu$ by appropriate momenta. One
 gets ~:
\bea &&h~ =~-2~\frac{\epsilon^{\mu\alpha\beta\delta}~J_\mu~ p_\alpha~
p_{1 \beta}~p_{2\delta}~}{s~[(t-m_B^2)~(q^2-t)~-~s ~t]} \\
 &&r~ =~2~i~\frac{~-2~s~p^\mu J_\mu~+2~(m_B^2-u-t)~(p_1+p_2)^\mu
 J_\mu~+~(t-u)~(p_1-p_2)^\mu
 J_\mu ~}{4~ s ~m_B^2~-~(2 m_B^2-u-t)^2~+~(u-t)^2~}\\
 &&w_-~ =~i~\frac{(p_1-p_2)^\mu
 J_\mu }{s}~-~\frac{t-u}{2 s}~r
\\
 &&w_+~ =~-i~\frac{(p_1+p_2)^\mu
 J_\mu }{s}~+~\left[1+\frac{t+u}{2 s} -\frac{m_B^2}{s} \right]~r
 \;\; . \label{ff4} \eea

 The calculation of the trace in
(\ref{ampiezza}) is straightforward and is similar to those
performed in  \cite{Brho} and  \cite{Bpi} for similar processes.
The evaluation of the integral is however  more complicated,
because  the kinematics  is more involved, due to the presence of
an extra momentum. The integration can be performed numerically,
but is time consuming, because, unlike the semileptonic
decays with one hadron in the final state, where the integration
involves one variable, here the integration domain is genuinely
three-dimensional.   The calculation becomes simpler putting the
light quark mass $m_q=0$ in the relevant formulae, which is an
approximation we perform and is justified by the small value of
$m_q$ in our model. Similarly we put $m_\pi=0$. Nevertheless the
computation remains huge, since each of the four form factors
depends on three variables and the number of points needed
to have a good accuracy is high.

An important point to be stressed is the kinematical range in
which the predictions of the present model are reliable. We cannot
pretend to extend our analysis to very small pion momenta for the
following reasons.  First of all, as discussed in \cite{Bpi}, when
$|\vec p_\pi|\to 0$ the results of the model become strongly
dependent on a numerical input of our calculation, i.e. the value
of the light quark mass $m_q$. The numerical value of $m_q$ cannot
be fixed adequately because the values of the quark masses were
fitted from the heavy meson spectrum, which is not very sensitive
to  $m_q$ ( for more details see \cite{pietroni} ). Therefore the
value we consider
 in the model  $m_q\simeq 38$ MeV (or $m_q = 0$ in the present
approximation) has considerable uncertainties. For large or
moderate $|\vec p_\pi|$ this uncertainty does not affect the
numerical results: the pion momenta are sufficiently large to
render the results insensitive to the actual value of $m_q$. For
very small  $|\vec p_\pi|$ the numerical results depend strongly
on $m_q$, which makes them unreliable. This is the first reason to
exclude the soft pion limit from the analysis also in this paper.
A second reason is that, in  the soft pion limit, the role of pole
diagrams such as those studied in \cite{lee} becomes relevant.
These diagrams cannot be accounted for by the  present scheme,
which at most can be used to model a continuum of states,
according to the quark-hadron duality ideas. The low-lying
resonances, such as those studied in \cite{lee} should be added
separately\footnote{This is the reason why in \cite{Bpi} the $B^*$
pole of the $B\to\pi$ form factor is not reproduced in the $|\vec
p_\pi|\to 0$ region.}. The same should be said about the
resonances encountered at small $s$, such as the $\rho$-resonance.
This resonance is not considered in \cite{lee}, but is expected to
play a major role; indeed experimentally one has $\dd {\cal
BR}(\bar B^0\to \rho^- \ell^+ \nu_\ell) = 2.5^{+0.8}_{-1.0}\times
10^{-4} $, which shows that this is a relevant piece of the
$B-$decay width into two pions. Therefore we  assume a lower
cutoff $s\geq~s_0$, with $s_0=1$ GeV$^2$ and we expect that the
results are not affected by the above-mentioned theoretical
uncertainties. Since the $\rho-$resonance and the chiral
contributions discussed in \cite{lee} are absent in our approach,
their contribution should be added separately. We expect a large
contribution from the $\rho$ and a tiny contribution from the
diagrams discussed in \cite{lee} since they are significant in a
very small region of the phase space (see the discussion in
\cite{lee}).

For $s \geq s_0$ our model has no similar limitations. By duality
we would expect that the sum over higher mass resonances can be
reproduced fairly well by the continuum model we employ here:
therefore these higher resonances should not be separately
considered to avoid double counting problems. It could be
observed, in this context, that the failure observed in \cite{Bpi}
at high $q^2$ (for the $B \to \pi$ semileptonic decay) would
correspond, in the present case, to the small $s$, not to the
large $s$ region.

\begin{figure}[ht!]
\begin{center}
    \includegraphics*[width=0.45\textwidth]{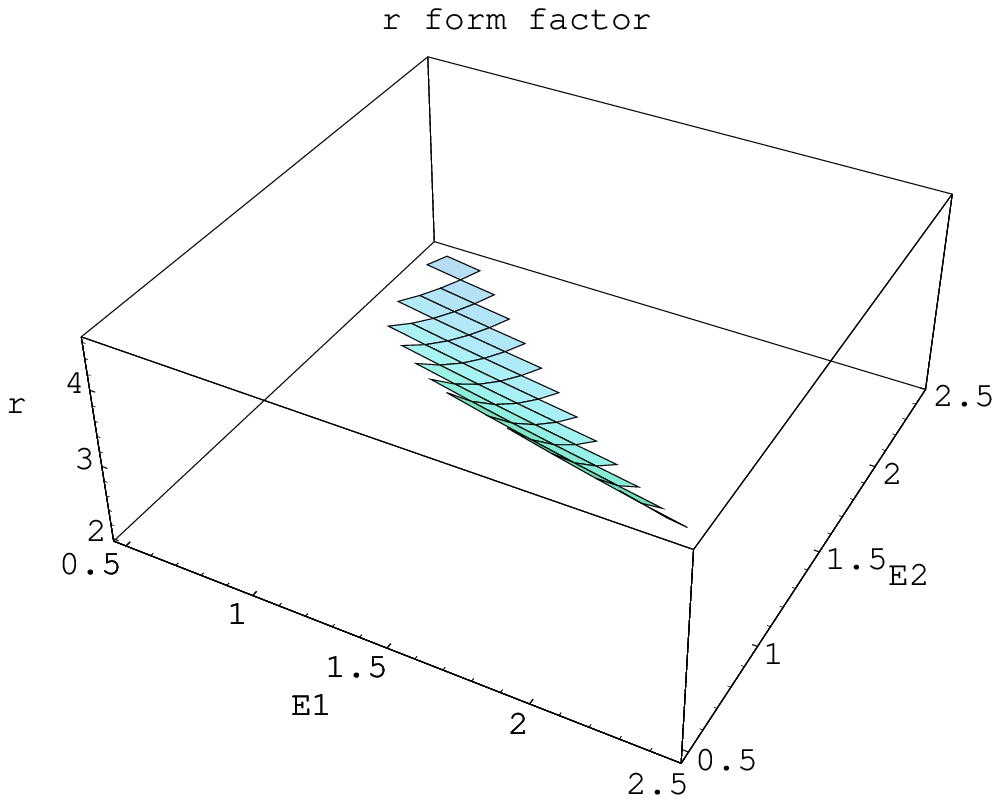}
    \includegraphics*[width=0.45\textwidth]{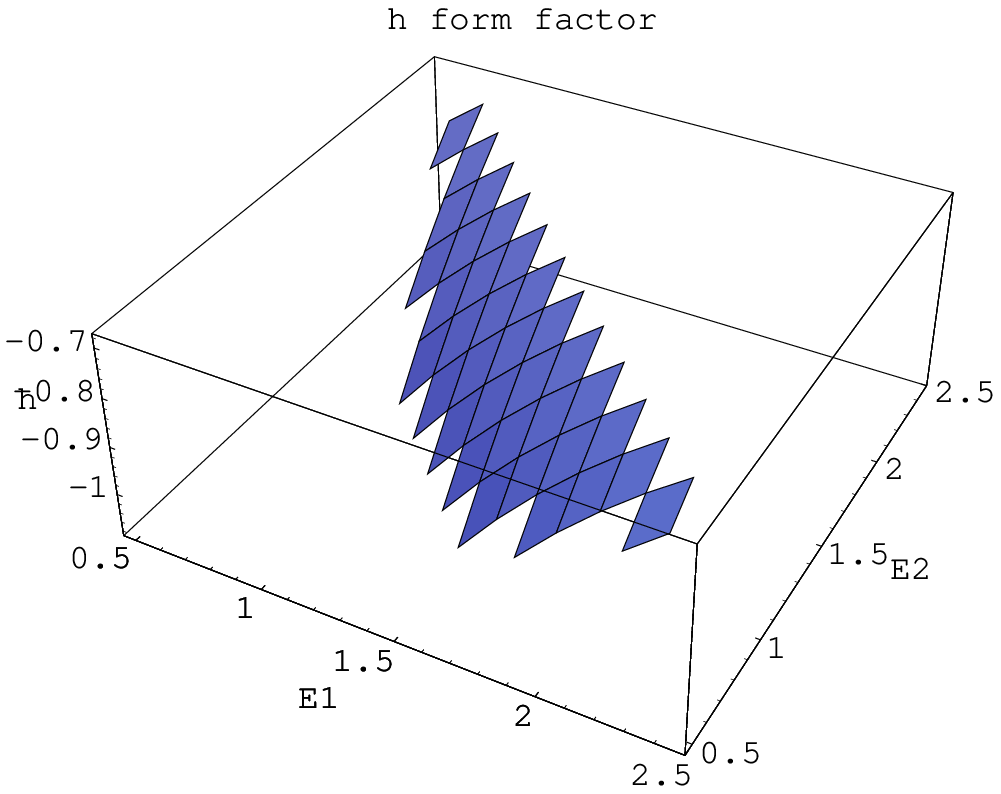}\\*
    (a)\hspace{0.5\textwidth}(b)
\end{center}
\begin{center}
    \includegraphics*[width=0.45\textwidth]{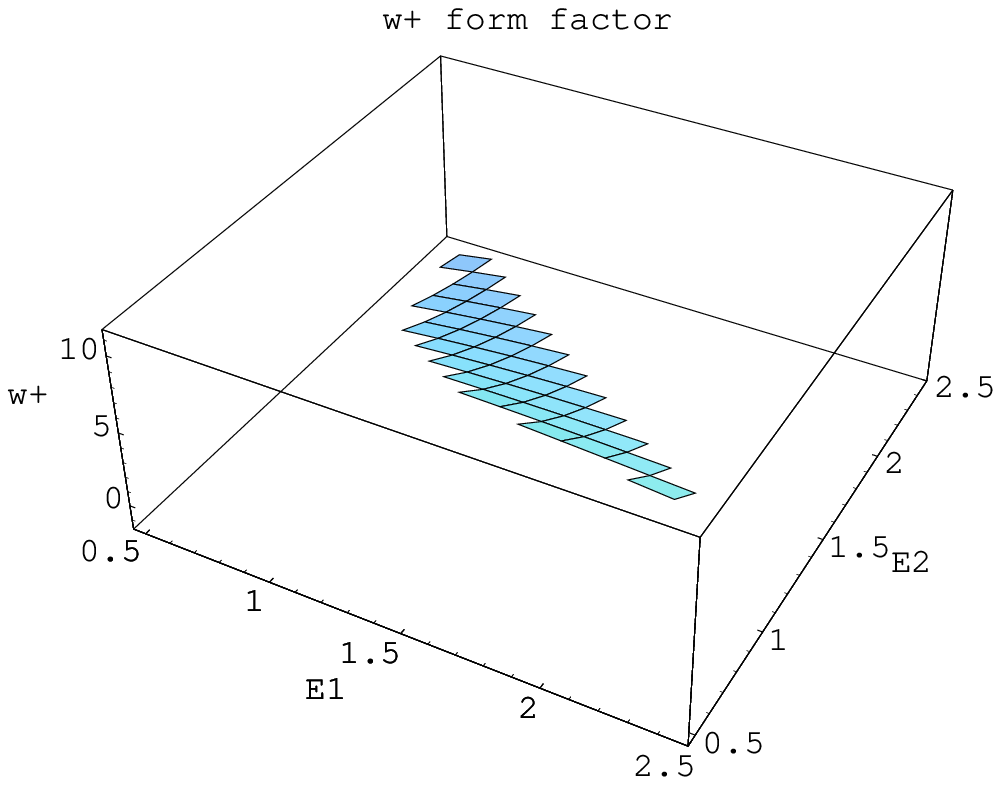}
    \includegraphics*[width=0.45\textwidth]{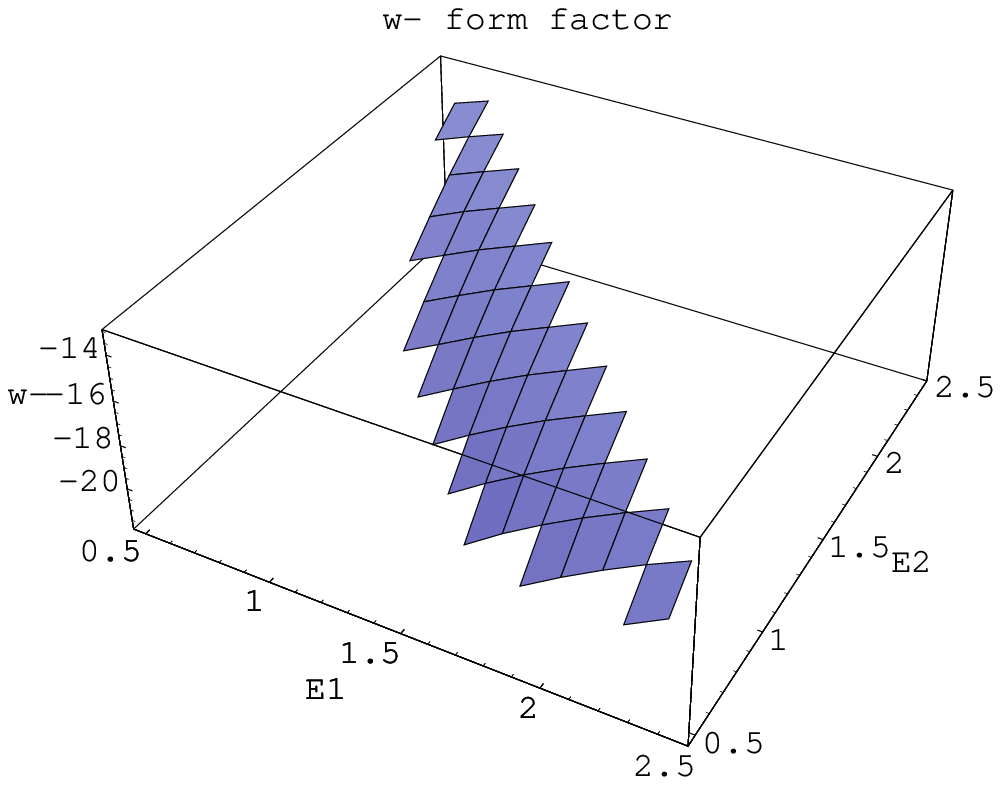}\\*
    (c)\hspace{0.5\textwidth}(d)
\end{center}
\caption{The relevant form factors for  $B^- \to \pi^+
\pi^-$ semileptonic decay at $s=5$ GeV$^2$ as a function of $E_1$
and $E_2$. Units are GeV$^{-1}$ for $w_+$ in (c), $w_-$ in (d) and
$r$ in (a), GeV$^{-3}$ for $h$ in (b).} \label{f:fig2}
\end{figure}

 Instead of presenting the form factors as functions of $s$, $q^2$
 and $t$ we prefer to consider $s$, $E_1$ and $E_2$, the pion
 energies.
In terms of $E_1$,  $E_2$ and $s$ the allowed kinematical range is
as follows (we put $m_\pi=0$):
\begin{eqnarray}\label{rel2}
s_0~\leq& s &\leq~m_B^2\nn\\ \frac{s}{2 m_B}~\leq & E_2 &
\leq~\frac{m_B}{2} \nn\\
\frac{s+m_B^2}{2m_B}-E_2~\leq& E_1 & \leq~\frac{s}{4 E_2}
\end{eqnarray}
 \vskip 0.2truecm
In tables 1-4 we present some numerical results for the form
factors $h(s,~E_1,E_2)$,  $r(s,~E_1,E_2)$,
$w_+(s,~E_1,E_2)$ and
$w_-(s,~E_1,E_2)$ in the $B^- \to \pi^+~ \pi^-$ semileptonic decay.
In each table we present all
the form factors at fixed $s$ and different values of the
$(E_1,~E_2)$ pair ($s=1$ GeV$^2$ in table 1, $s=5$ GeV$^2$ in
table 2, $s=10$ GeV$^2$ in table 3, $s=19$ GeV$^2$ in table 4).
These results should allow to
get a quantitative assessment of the numerical relevance of the
various form factors in the allowed kinematical range. A graphical
presentation of the fitted numerical output for $s=5$ GeV$^2$
is given in
figs. \ref{f:fig2} for the four form factors.
 A different way to present the data is  to introduce averaged
form factors. We choose to perform an average in the pion energies
according to the following formula:
\begin{equation}\label{averaged}
f(s)=\frac{1}{\Delta (s)}  \int_{s/(2 m_B)}^{m_B/2} d E_2~
\int_{s/(4E_2)}^{(s+m_B^2)/(2m_B)-E_2}
d E_1~f(s,~E_1,E_2)~,
\end{equation}
valid for all the form factors ($f=h,~r,~w_+,~w_-$). Here $\Delta
(s)$ is the allowed area in the $(E_1,~E_2)$ plane:
\begin{equation}\label{area}
\Delta (s)=\frac{m_B^4-s^2}{8
m_B^2}+\frac{s}{4}\ln\frac{s}{m_B^2}~.
\end{equation} The numerical results we obtain have an average error
around $10 \%$.   A simple way to present the data is by an
analytical formula: for example the data can be fitted by the
following relation
\begin{equation}\label{ave}
f(s)~=~\frac{\beta~(s-s_1)~(s-s_2)}{(s-s_3)^2~+~s_4^2}~,
\end{equation}a procedure which introduces an average
numerical error of $\pm 10\%$; we stress, however that in
computing the width we have not used this fit and therefore this
further error has not been introduced. We also point out that the
Breit-Wigner shape of eq. (\ref{ave}) is a useful parameterization
and has no dynamical meaning.

\begin{figure}[ht!]
\begin{center}
    \includegraphics*[width=0.48\textwidth]{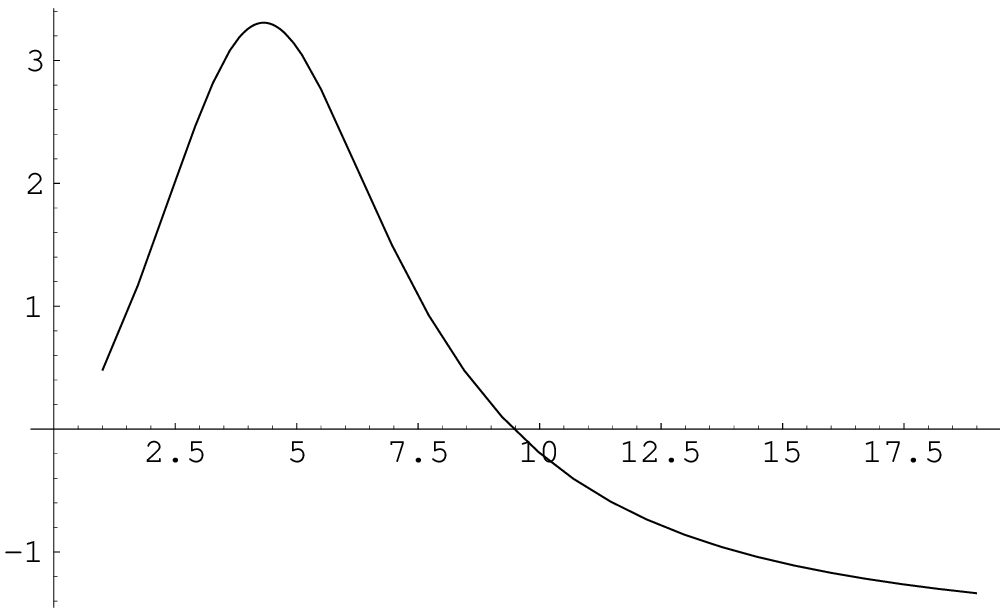}
    \includegraphics*[width=0.48\textwidth]{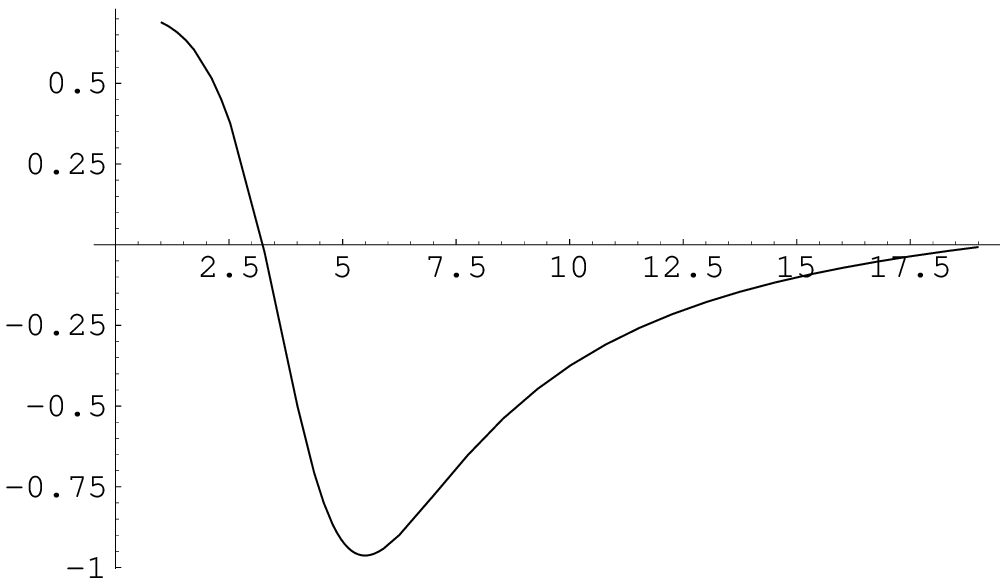}\\*
    (a)\hspace{0.5\textwidth}(b)
\end{center}
\begin{center}
    \includegraphics*[width=0.48\textwidth]{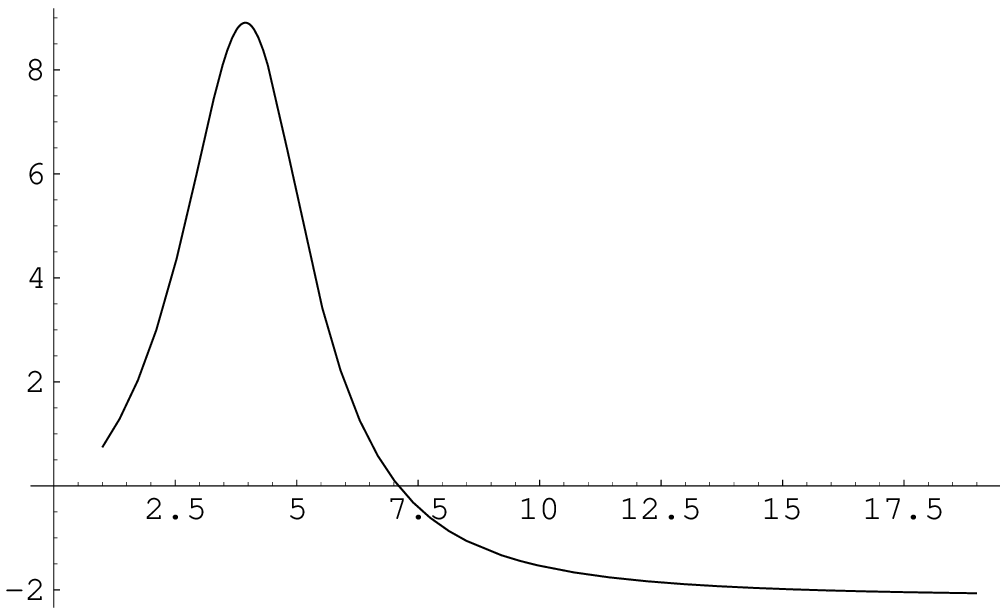}
    \includegraphics*[width=0.48\textwidth]{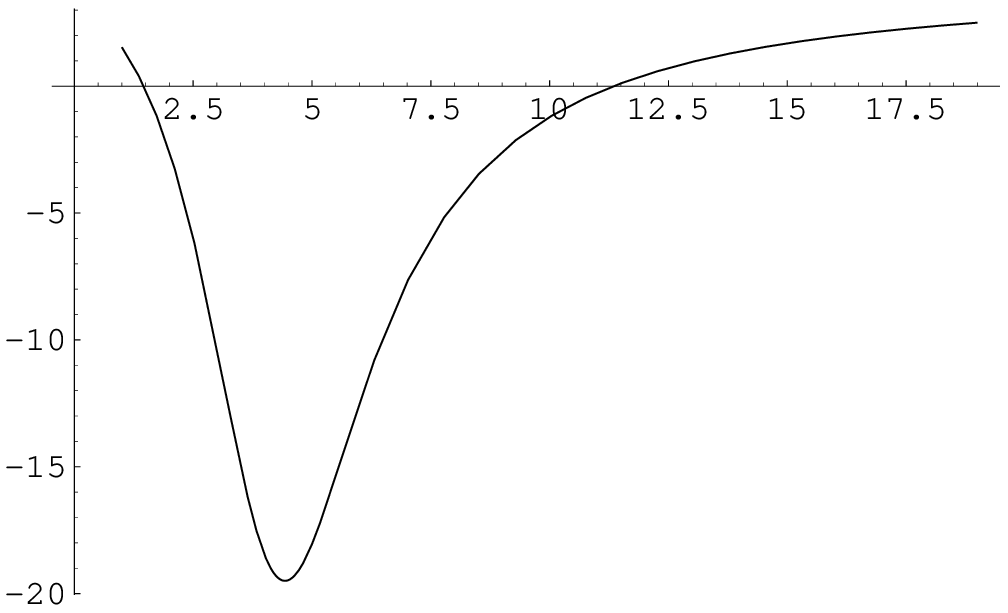}\\*
    (c)\hspace{0.5\textwidth}(d)
\end{center}
\caption{The averaged form factors for  $B^- \to \pi^+
\pi^-$ semileptonic decay. Units are GeV$^{-1}$ for $w_+$ in (c),
$w_-$ in (d) and $r$ in (a), GeV$^{-3}$ for $h$ in (b).} \label{f:fig6}
\end{figure}

 The values of the
coefficients $s_k~,\beta$ appearing in (\ref{ave}) are reported in
table 5  for all the form factors of the $B^-$ decay. The form
factors are depicted in figs. \ref{f:fig6}. \vskip 0.2truecm We
observe that due to the limitations of our approach, the
kinematical region of validity of the present model has no overlap
with the soft pion region where pole diagrams, see e.g.
\cite{lee}, are expected to dominate. Therefore
 a comparison
of our work with the results of these  models   is impossible.
\par
\noindent
 Let us now evaluate the partial width $\Gamma(B \to \pi\pi \ell \nu)$.
 The relevant formulae to compute the width are
reported in \cite{lee} and we do not reproduce them here. We only
give our numerical results for the cut-off width ($s\geq 1 {\rm
GeV}^2$). Numerically we get
\be
{\cal BR}( B^-\to \pi^+\pi^- \ell^- \bar \nu_\ell) = 2.2 \
\left(\frac{|V_{ub}|}{3.2\times 10^{-3}}\right)^2\times 10^{-4}~{\hskip 1
truecm}~(s\geq1 {\rm GeV}^2)~. \ee  For the other decay channel we
have:
\be
{\cal BR}(\bar B^0\to \pi^+\pi^0 \ell^- \bar \nu_\ell) = 3.2 \
\left(\frac{|V_{ub}|}{3.2\times 10^{-3}}\right)^2\times 10^{-4}~~{\hskip 1
truecm}~(s\geq1 {\rm GeV}^2)~. \ee

The contribution of the $\rho$ resonance to these decay modes can
be estimated in the present model \cite{Brho} as follows:
\be
{\cal BR}( B^-\to \pi^+\pi^- \ell^- \bar \nu_\ell)\Big|_\rho = 1.2
\ \left(\frac{|V_{ub}|}{3.2\times 10^{-3}}\right)^2\times
10^{-4}~. \ee  For
the other decay channel we have:
\be
{\cal BR}(\bar B^0\to \pi^+\pi^0 \ell^- \bar \nu_\ell)\Big|_\rho = 2.4
\ \left(\frac{|V_{ub}|}{3.2\times 10^{-3}}\right)^2\times 10^{-4}~. \ee
This latter branching ratio is in agreement with the
experimental figure quoted above.

We can therefore conclude that from an experimental point of view
the semileptonic decay channel with two non-resonant pions in the
final state represents an interesting process with a significant
branching ratio, of the same order of magnitude of the single pion
or the single $\rho$ semileptonic decay mode. It would be nice to
find this decay mode in the future experimental analysis and to
test the present prediction of the QCD relativistic potential
model.  \vskip 0.2cm
\par\noindent{\bf Acknowledgements.} We wish to thank P. Colangelo,
F. De Fazio, M. Pellicoro and A. D. Polosa for useful comments.

\newpage

\begin{table}
\caption{Numerical values of the   form factors  for several
values of $E_1~$,  $~E_2$  (in GeV)  and $s~=~1$ GeV$^2$. Units
are GeV$^{-1}~$ ($r$, $w_+$ and  $w_-$~) and GeV$^{-3}~$ ($h$).}
\begin{center}
\begin{tabular}{|c|c|c|c|c|}
  \hline
 $ (E_1,~E_2)~ $ &~$r$ & $h$ & $w_+$ & $w_-$   \\ \hline
   $~ (0.14~,~2.36) ~$ &~$-~0.26 $ ~&~ 0.53 ~&~ 5.99 ~&~ 3.12
\\ $~ (0.18~,~2.07) ~$ &~$-~0.23 $ ~&~ 0.62 ~&~ 5.89 ~&~ 2.83
\\ $~ (0.23~,~1.79) ~$ &~$-~0.16 $ ~&~ 0.73 ~&~ 5.89 ~&~ 2.44
\\ $~ (0.28~,~1.51) ~$ &~$-~0.039$ ~&~ 0.88 ~&~ 5.77 ~&~ 1.92
\\ $~ (0.35~,~1.23) ~$ &~$~ 0.18 $ ~&~ 1.07 ~&~ 5.60 ~&~ 1.28
\\ $~ (0.43~,~0.94) ~$ &~$~ 0.56 $ ~&~ 1.32 ~&~ 5.41 ~&~ 0.56
\\ $~ (0.57~,~0.66) ~$ &~$~ 1.22 $ ~&~ 1.60 ~&~ 5.18 ~&-~0.019
\\ $~ (0.85~,~0.38) ~$ &~$~ 2.21 $ ~&~ 1.72 ~&~ 4.20 ~&~ 0.58
\\
\hline
\end{tabular}
\end{center}
\end{table}


\begin{table}
\caption{Numerical values of the   form factors  for several
values of $E_1~$,  $~E_2$  (in GeV)  and $s=5$ GeV$^2$. Units are
GeV$^{-1}~$ ($r$, $w_+$ and  $w_-$~) and  GeV$^{-3}~$ ($h$).}
\begin{center}
\begin{tabular}{|c|c|c|c|c|}
  \hline
 $ (E_1,~E_2)~ $ &~$r$ & $h$ & $w_+$ & $w_-$   \\ \hline
   $~ (0.54~,~2.40) $ ~&~ 2.15    ~&~ -~0.76  ~&-~1.25 ~&-~13.9
\\ $~ (0.62~,~2.16) $ ~&~ 2.36    ~&~ -~0.84  ~&-~0.31 ~&-~15.1
\\ $~ (0.71~,~1.92) $ ~&~ 2.60    ~&~ -~0.92  ~&~ 0.85 ~&-~16.5
\\ $~ (0.82~,~1.68) $ ~&~ 2.88    ~&~ -~0.99  ~&~ 2.27 ~&-~17.9
\\ $~ (0.96~,~1.44) $ ~&~ 3.22    ~&~ -~1.06  ~&~ 3.98 ~&-~19.3
\\ $~ (1.14~,~1.20) $ ~&~ 3.62    ~&~ -~1.10  ~&~ 6.01 ~&-~20.7
\\ $~ (1.40~,~0.95) $ ~&~ 4.08    ~&~ -~1.08  ~&~ 8.29 ~&-~21.5
\\ $~ (1.82~,~0.71) $ ~&~ 4.52    ~&~ -~0.95  ~&~ 10.5 ~&-~20.9
\\
\hline
\end{tabular}
\end{center}
\end{table}


\begin{table}
\caption{Numerical values of the   form factors for several values
of $E_1~$,  $~E_2$  (in GeV)  and $s=10$ GeV$^2$. Units are
GeV$^{-1}~$ ($r$, $w_+$ and  $w_-$~) and GeV$^{-3}~$ ($h$).}
\begin{center}
\begin{tabular}{|c|c|c|c|c|}
  \hline
 $ (E_1,~E_2)~ $ &~$r$ & $h$ & $w_+$ & $w_-$   \\ \hline
   $~ (1.03~,~2.45) ~$ &~ ~ 0.42  ~&-~0.29 ~&-~2.71 ~&-~0.36
\\ $~ (1.13~,~2.26) ~$ &~ ~ 0.29  ~&-~0.32 ~&-~2.71 ~&-~0.33
\\ $~ (1.24~,~2.08) ~$ &~ ~ 0.14  ~&-~0.34 ~&-~2.68 ~&-~0.37
\\ $~ (1.37~,~1.89) ~$ &~ -~0.044 ~&-~0.37 ~&-~2.60 ~&-~0.49
\\ $~ (1.52~,~1.70) ~$ &~ -~0.24  ~&-~0.40 ~&-~2.42 ~&-~0.75
\\ $~ (1.70~,~1.51) ~$ &~ -~0.45  ~&-~0.42 ~&-~2.11 ~&-~1.15
\\ $~ (1.93~,~1.32) ~$ &~ -~0.63  ~&-~0.44 ~&-~1.60 ~&-~1.78
\\ $~ (2.23~,~1.13) ~$ &~ -~0.75  ~&-~0.44 ~&-~0.81 ~&-~2.72
\\
\hline
\end{tabular}
\end{center}
\end{table}

\begin{table}
\caption{Numerical values of the   form factors for several values
of $E_1~$,  $~E_2$  (in GeV)  and $s=19$ GeV$^2$. Units are
GeV$^{-1}~$ ($r$, $w_+$ and  $w_-$~) and GeV$^{-3}~$ ($h$).}
\begin{center}
\begin{tabular}{|c|c|c|c|c|}
  \hline
 $ (E_1,~E_2)~ $ &~$r$ & $h$ & $w_+$ & $w_-$   \\ \hline
   $~ (1.87~,~2.55) ~$ &~ -~0.81  ~&~  0.0041 ~&-~1.31 ~&~ 1.73
\\ $~ (1.94~,~2.45) ~$ &~ -~0.87  ~&~  0.0028 ~&-~1.37 ~&~ 1.77
\\ $~ (2.02~,~2.36) ~$ &~ -~0.95  ~&~  0.0012 ~&-~1.44 ~&~ 1.80
\\ $~ (2.10~,~2.27) ~$ &~ -~1.02  ~&~  0.00092~&-~1.51 ~&~ 1.82
\\ $~ (2.19~,~2.17) ~$ &~ -~1.10  ~&-~ 0.0033 ~&-~1.57 ~&~ 1.83
\\ $~ (2.29~,~2.08) ~$ &~ -~1.18  ~&-~ 0.0062 ~&-~1.64 ~&~ 1.83
\\ $~ (2.40~,~1.99) ~$ &~ -~1.27  ~&-~ 0.0096 ~&-~1.70 ~&~ 1.81
\\ $~ (2.51~,~1.89) ~$ &~ -~1.35  ~&-~ 0.014  ~&-~1.75 ~&~ 1.77
\\
\hline
\end{tabular}
\end{center}
\end{table}

\begin{table}
\caption{Numerical values of the parameters appearing in the
formula
$\dd{f(s)~=~\frac{\beta~(s-s_1)~(s-s_2)}{(s-s_3)^2~+~s_4^2}}~$.
$f(s)$ is any of the averaged form factors.~ Units are GeV$^{2}~$
for $s_1~$, $s_2~$, $s_3~$ and $s_4$; $f(s)$ and $\beta$ have the
same units.}
\begin{center}
\begin{tabular}{|c|c|c|c|c|c|}
  \hline
 {\rm form\ factor} & $s_1$ & $s_2$ & $s_3$ & $s_4^2$ & $\beta$ \\ \hline
 $~ w_+~$ & $-~0.957$ & $+~7.15$ & $+~3.98$ & $+~3.38$ & $-~1.87$ \\
 $~ w_-~$ & $+~1.77$  & $+~11.4$ & $+~4.20$ & $+~3.47$ & $+~3.91$ \\
 $~ h~$   & $+~3.25$  & $+~19.5$ & $+~5.38$ & $+~7.93$ & $+~0.207$ \\
 $~ r~$   & $-~0.446$ & $+~9.47$ & $+~4.32$ & $+~11.5$ & $-~1.57$  \\
\hline
\end{tabular}
\end{center}
\end{table}

\end{document}